# TOWARDS HIGH PERFORMANCE COMPUTING (HPC) THROUGH PARALLEL PROGRAMMING PARADIGMS AND THEIR PRINCIPLES

Dr. Brijender Kahanwal

Department of Computer Science & Engineering, Galaxy Global Group of Institutions, Dinarpur, Ambala, Haryana, India

## ABSTRACT

*Nowadays, we are to find out solutions to huge computing problems very rapidly. It brings the idea of parallel computing in which several machines or processors work cooperatively for computational tasks. In the past decades, there are a lot of variations in perceiving the importance of parallelism in computing machines. And it is observed that the parallel computing is a superior solution to many of the computing limitations like speed and density; non-recurring and high cost; and power consumption and heat dissipation etc. The commercial multiprocessors have emerged with lower prices than the mainframe machines and supercomputers machines. In this article the high performance computing (HPC) through parallel programming paradigms (PPPs) are discussed with their constructs and design approaches.*

## 1. INTRODUCTION

The numerous computational concentrated tasks of the computer science like weather forecast, climate research, the exploration of oil and gas, molecular modelling, quantum mechanics, and physical simulations are performed by the supercomputers as well as mainframe computer. But these days due to the advancements in the technology multiprocessors systems or multi-core processor systems are going to be resembled to perform such type of computations which are performed by the supercomputing machines.

Due to the recent advances in the hardware technologies, we are leaving the von Neumann computation model and adopting the distributed computing models which have peer-to-peer (P2P), cluster, cloud, grid, and jungle computing models in it [1]. All these models are used to achieve the parallelism and are high performance computing (HPC) models.

*Concurrency and Parallelism:* The terms concurrency and parallelism must be clear in our minds first. It can be well explained with the help of the threads (light weight processes). When two or more threads are in the middle of execution process at the same time, actually, they may or may not be executing at the same time, but they are in the middle of it.

   

International Journal of Programming Languages and Applications ( IJPLA ) Vol.4, No.1, January 2014

It is called concurrency [25]. The threads may or may not execute at the single processor or a multiprocessor machine in it. When two or more threads are actually running at the same time on different CPUs, it is known as parallelism [25]. To achieve parallelism, we always require at least two CPUs which may be on a single machine (multiprocessor machine) or more machines. The parallel events may also be called as concurrent events, but the reverse is not true always. It is well described with the help of set theory that parallelism ⊂ concurrency (Parallelism is contained in Concurrency) as shown in Fig.1.

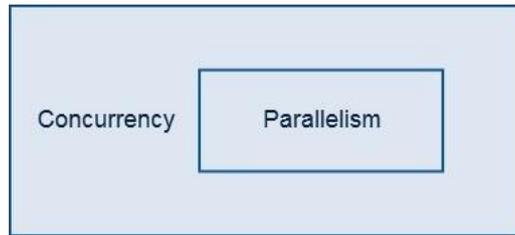

Figure 1: Concurrency is the superset of Parallelism

On the Von Neumann computing machines, the programming is a single execution sequence. But there might be various subroutines that can be executed simultaneously within a single program. These are called sequential due to the execution of subroutines proceeds in predetermined sequence. In general cases, the programs are termed as concurrent or parallel in which the subroutines can be executed concurrently and these subroutines are known as tasks [2].

Now a day, it is a common practice now to execute various programs concurrently by the computing machines. It may have the architecture with multiprocessors (various CPUs) which share the common memory space as shown in the Fig.2 (a) or another architecture that may have multiprocessors with their independent memories or distributed memories as shown in the Fig.2 (b).

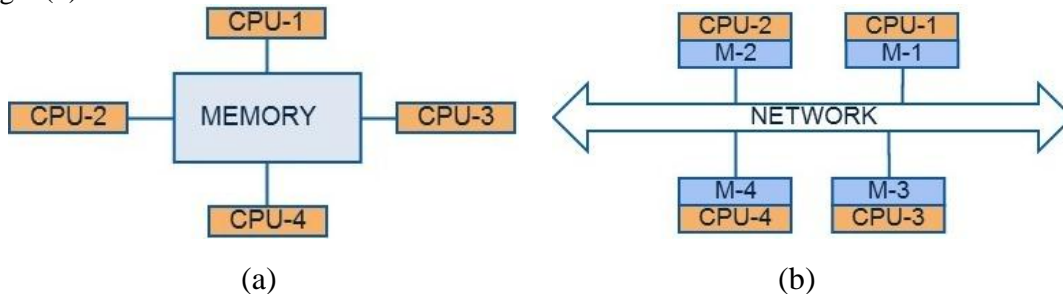

                    (a)                     (b)

Figure 2 (a): Shared Architecture.         Figure 2 (b): Distributed Architecture.

It is the big challenge for the scientists to utilize these hardware technologies efficiently, effectively, and these processors may work cooperatively. In the present scenario, software technologies (STs) are not having well compatibility with the hardware technology's growth that the STs can utilize them efficiently and effectively. Hence the parallel computing community is going to be aware that they can build software technologies which are efficient and effective. But till now the programmers as well as the scientists are incompetent to find the solution. So it is the need to get more awareness regarding the parallelism, so we can find the better solutions for high performance computing (HPC). The article contains more sections which are organized as follows: the related works will be available in the section 2.

## 2. RELATED WORKS

The parallelism is not a novel concept of computing. The Law of Amdahl is the key principle to estimate maximum improvements in the components of the system [3] which brings the idea of parallel computing to find the optimum performance. The time 1960-70 was the boom time for the parallel computing and during this time, we have solved many problems in achieving the optimum performance, but they have encountered today. Multi-core chips are a new paradigm in parallelism. Parallel computation is the never-ending desire for much faster and much cheaper computation of level of supercomputers as well as mainframe computers [4]. Until now we have not remarkable progress in building the efficient and optimal softwares for utilizing the parallel computer architectures of today [5].

## 3. PARALLEL PROGRAMMING CONSTRUCTS OR PRINCIPLES

It is complicated to write the parallel programs as compared to write sequential programs. We design algorithms and express them in some programming languages to execute on the computing machines. In the case of parallel programming we have to develop the same functioning, but it also adds more challenges to it. Such types of challenges are as follows: structured constructs[6]: structured region; thread based constructs[7]: synchronization, critical sections, and deadlock; and object-oriented constructs [8]: object replication, latency hiding, termination detection, and user-level scheduling; concurrency; data distribution; inter-process communication; computational load balancing; variable definitions [2]; parallel compositions [2]; program structures [2]; and easy implementation and debugging;. All of these are explored in the following sub sections.

### 3.1. Structured Construct – Structured Region

The structured parallel programming construct is introduced as a structured region. It has a region name and a region body which is enclosed with two barriers namely entry barrier as well as exit barrier. A par (or parfor) block has the starting instruction as the region name. A region name has a region keyword then an arbitrary name given to the region by the programmer and a list of the participants (processes) [6]. If the participants list is declared explicitly then the specified names becomes the participants and if the list is not mentioned then all the processes are the participants.

This structured region semantics is very easy and clear. The entry in the region is done only if all the participants reach at their specific entry points. There are the unique effects of the execution of the region body. The region is exited by all the participants after the complete execution of all the operations of the region body. A structured region has a single entry and exit point [6]. It wraps up inter-process communication as well as synchronization operations in it and makes the parallel programming easier to understand and less error-prone. It is opposite to the concept of mutual exclusion in which only a single process can enter in the critical region. But in the case hare all the processes can enter in the structured region.

### 3.2. Thread-Based Constructs

The process and the thread are much related terms. A process is program in execution and there may be some independent units within a process which are known as the threads. A thread is dispatch-able work unit. It is also known as the light-weight process. So it is concluded that the

International Journal of Programming Languages and Applications ( IJPLA ) Vol.4, No.1, January 2014

threads makes a process or it is a subset of a process [10]. Both the process as well as the thread is an active entity and a simple program before execution is a passive entity. The threads of a single process share same address space, so the context-switching as well as communication between threads is inexpensive [10]. The sharing between the threads creates some difficulties which are explored in the following sub-sections.

1. *Synchronization:* It is the construct which enforces the mechanism for controlling the execution order of the threads and resolves the conflicts among the threads [7]. It is a way of coordinating the execution of the threads and managing the shared address space.
   In synchronization, mutual exclusion and condition synchronization operations are used widely. In mutual exclusion, one of the threads block the critical section (shared data area by the threads) and other threads will wait for getting their turns one by one. The scheduler controls for the turns. But in the case of conditional synchronization, threads are blocked until some particular condition is satisfied. Here the thread has to wait until a particular condition is achieved. So the synchronization is well managed by the programmer or by the programming system, it is a critical construct for multi-threaded programming.
2. *Critical sections:* These sections have shared dependency variables and many threads are dependent on them [7]. It is the great programming construct for thread-based programming, so the threads can use these sections mutually exclusively and prevent to use these sections simultaneously. These sections should be minimized in size.
3. *Deadlock:* It is the situation when a thread holds a lock and waiting for another lock which is held by another thread and this thread is waiting for the lock first to be released. Such as the code: T1: lock (1); lock (2); and T2: lock (2); lock (1); in this code, the deadlock may or may not occur. The four basic conditions need to be hold which are mutual exclusion; hold-and-wait; no pre-emption; and circular wait.

### 3.3. Object-Oriented Constructs

The object-oriented parallel programming has complex computational as well as communicational structures to achieve the efficiency or optimization. For improving the performance in the object-oriented programming languages some of the constructs are discussed in the following sub-sections [8].

1. *Object Replication:* This construct highly improves the performance in the distributed memory architectures. When a program is frequently accessing an object then it is better to create a local replica of it for the processor and then there is a big fall in the number of remote messages [8].
2. *Latency Hiding:* It is an optimization technique which reduces the waiting time for the remote messages. Here local computations and remote communications are overlapped. In it we break up a single thread into multiple threads manually by modifying the program [8].
3. *Termination Detection:* In few parallel applications like search problems, it is a typical task to detect the termination point because of the invoking of the many threads and finding their termination points in absence of the global control on them [8].
4. *User-level Scheduling:* A proper scheduling at application level also improves the performance of the parallelism. User-level scheduling facility is not offered in most of the programming language systems, so it becomes necessary to provide it by the programmers explicitly to control the order of execution [8].

### 3.4. Concurrency

These days, the processors are inexpensive as compared to the previous time, so we are constructing the distributed systems. Due to such things in the development the concurrency has a little importance. The programmers are working on various types of applications like DBMS, L-S parallel technical computations, real time applications, and embedded systems etc [9].
When a concurrent program shares one or more processors during execution is known as multiprogramming; when its sub-processes are to be executed on independent processor then it is known as multiprocessing; when there is the addition of the communication network then it is known as distributed processing; and any combination of these is known as hybrid approach [9]. Still, it is the fundamental construct to utilize optimally the parallel computing resources. We can't achieve the parallelism without dividing the operations to execute concurrently. A problem has many sub-problems in it for concurrent execution; there is a need to differentiate the concurrent tasks within the main problem. It is the dexterity of the programmers.

### 3.5. Data Distribution

It is a big challenge to distribute the data which creates problem. In the parallelism there are so many processors which are working cooperatively. Now a day, the principle of locality is important for the better performance of the systems. But in the case of parallelism it becomes the problem or a decision making event which data to be localized for the particular processor. It is due to the concept of independent cache memories for each processor in the shared memory systems. For the parallel programmers, it becomes issue to manage it carefully. The performance of the system increases as we store more data in the caches because the processor can access it quickly as compared to the shared memory area.

### 3.6. Inter-process Communication

When we are going to execute a process on two or more processors, it becomes necessary to make communication among them for transferring data from one processor's cache memory to another processor's cache memory. So here is the need of maintaining caches of the processors with the mechanism called cache coherence that may be implemented via hardware or cache coherence protocols. Another case may be that the processors may have distributed memories and all the processors need to be communicated properly. There may be the need of explicit calls to a library which require transferring values among processors. There may be the communication overheads which must be minimized to get the advantage of the parallelism.

### 3.7. Computational Load Balancing

In the parallelism, there are two or more processors or separate machines which are connected through the network, to take the advantage of the parallelism all the processors or machines must be utilized properly and equally. The total computation must be equally distributed among the processors or machines for getting the benefits of high performance computing.

### 3.8. Variable Definitions

Two types of variables may be used in the programming languages namely mutable and definitional. The mutable variables are the normal variables which are used in the sequential programming languages. The assignment may be done to the variables and that may change during the program execution. The definitional variables are those variables in which we can assign values only once and they can be accessed by any number of tasks. In such variables there is no need to maintain synchronization.

### 3.9. Parallel Compositions

In the execution process, the statements are executed one after another and they also have additional sequential as well as conditional statements in the sequential programming language. To achieve the parallelism, the parallel statements must be added which becomes the additional threads of control to start the execution.

### 3.10. Program Structures

There may be two types of parallel program execution models. *Firstly*, transformational in which the main task is to transform the input data into the correct output value. *Secondly*, reactive or responsive in which the programs works regarding the events which are the external one.

### 3.11. Ease of Programming and Debugging

This is the issue for every type of programming language. The parallel programs must be easily implemented by the programmers. They do not require thinking more about the parallelism. The parallelism should be tackled by the programming language platforms. It is common to be bugs in the program implementation and there are so many side-effects of these bugs. So these may be removed easily with the help of good debugging tools.

## 4. PARALLEL PROGRAMMING APPROACHES

There are basically three approaches to program high performance computers (parallel computers). These are as follows:

### 4.1. Implicit Parallelism

It is also known as automatic parallelism. This approach is headache free for the programmer's point of view; here the complete working is done by the compilers to make parallel all of the executions [11]. All the parallel language constructs are inherently implemented by the language platform. Such type of job is always done in the pure functional programming languages. With the help of this approach the existing code is utilized on parallel systems. No changing is required in the existing code. It saves the development costs. And it is attractive for the vendors of the high performance computing.

Such type of parallelism has its own advantages as well as drawbacks. The advantages are as follows: *Firstly*, programmer's attention is completely on the algorithms. *Secondly*, we require very less code for programming. *Thirdly*, the productivity of the programmers increases as he/she does not care about the parallel programming constructs. *Fourthly*, the definitions of the algorithms are separated from the parallel executions. Fifthly, the legacy systems are utilized properly and which is the concept of re-usability.

The drawbacks are as follows: *Firstly*, the complete parallelism is not achieved because the programmers have much more information of the parallel potential (not efficient). *Secondly*, the programmers have not the exact control over the parallelism. *Thirdly*, there is no optimum parallel efficiency achieved. *Fourthly*, the algorithms which are already implemented may be executed with a low configuration system (architecture + memory) a few or more decade ago. But the recent configurations are very high with more storage capacity and processor speeds. *Fifthly*, it is a tough task for the scientists and researchers to design the parallel compilers.

### 4.2. Explicit Parallelism

In it the existing programming languages are utilized. The proper extensions are made to them to achieve all the parallel programming constructs [11]. Here the parallel programming principles are defined explicitly by the programmers. Explicit threading is a sub-approach of explicit parallelism in which the programmers creates parallel threads explicitly [22, 23].The explicit parallelism also has its own advantages and disadvantages.

The advantages are as follows: *Firstly*, the programmers are already trained in the existing language. *Secondly*, it is totally under the understanding and control of the programmers. The disadvantages are as follows: *Firstly*, it is very hard to debug and difficult to program for the programmers because everything is dependent on the creativity and thinking the programmers. *Secondly*, there is no standardization because there are so many extensions have been made by the developers with same functionality with different look.

### 4.3. Hybrid Parallelism

It is the mixed up approach which combines the features of implicit as well as explicit parallelism. It will take the advantages of both the above mentioned technique.

It is summarized that the language designers may design completely the new programming language paradigms which have all the parallel programming principles or constructs in it.

## 5. PARALLEL PROGRAMMING PARADIGMS

There are too many paradigms available for utilizing the current machines with parallel architectures. Some of the parallel programming languages are as follows:

### 5.1. Message Passing Interface (MPI)

It is a specification for message passing. It is de facto standard for the development of high performance computing applications for the distributed systems (heterogeneous networks) as well as parallel computers and clusters [12]. It has bindings for C, C++ and FORTRAN programming languages. It is highly portable environment. The workload partitioning as well as the work mapping are done explicitly be the programmers like Pthread [25] and UPC. All the communications between the processes take place with the help of message passing paradigm. In it one process sends the data to another process through message passing.

### 5.2. Fortress

It is also a thread-based specification programming language to design the HPC applications [12]. The work management, workload portioning, as well as work mapping may be done implicitly by the compiler as well as explicitly by the programmers. All for loops are parallel by

International Journal of Programming Languages and Applications ( IJPLA ) Vol.4, No.1, January 2014

default as implicit approach. The synchronization principles like reductions as well as atomic expression are specified by the programmers when there is a data competition in a program.

### 5.3. POSIX Threads (Pthreads) Programming

It is actually a set of C language types as well as procedure calls and all these are maintained or defined in a library named as pthread.h [12]. It is the duty of the programmers to maintain the shared data among the threads for avoiding the deadlocks and data races [25]. The pthread's create function has four parameters the task run thread, attribute, tasks to run in routine call, and routine argument. All has been closed with the help of pthread's exit function call. The workload partitioning and work mapping is done explicitly by the programmers.

### 5.4. OpenMP

It is also thread based open specification for shared memory architectures. It provides compiler directives, callable runtime library, and environment variables which extends the existing programming languages C, C++, and FORTRAN. It is portable platform [12]. The worker management is done impliedly and a little programmer's effort is required for the workload partitioning and task mappings, they are also performed implicitly. Programmers are required to tell the parallel region with the help of the compiler directives. The synchronization is also maintained implicitly by the OpenMP.

### 5.5. CILK (pronunciation as '*silk'*)

It is a multi-threaded programming language. It is appropriate for the recent multi-core CPU architectures. It is based on the traditional programming language C. Cilk a true parallel extension to C semantically with good performance [13]. In 1994, it was designed by the MIT scientists. In it the work-stealing scheduler is efficiently utilized. A Cilk program is a collection of Cilk procedures and every procedure has a sequence of threads. Every thread is non-blocking C language function which can run independently without waiting or suspension.

### 5.6. OpenMPI

It is the programming tool which is specially designed for the poor scientific programmers for achieving simple and routine parallelism. It is based on the existing programming tool OpenMP [14]. It provides the sufficient directives for achieving the parallelism. All the directives are followed by the notation directive *pragma ompi*. The few of the directives are distvar (dim=dimension, sleeve=s size) for the distributed array on parallel processes; global for declaring the variable as global variable; for (reduction (operator: variable)) to parallelize the for loop; syn sleeve (var=variable list) for exchanging the sleeve data of the distributed array for correctness ; sync var (var=variable list, master=node id) for synchronizing the global variable by coping the master data to others; and single (master=node id) for executing the next block by one process only as a delegate for other processes.

### 5.7. JAVA

It is the most popular programming language these days because we can create common applications on it and it also supports the parallelism through its multi-threading concept. It uses the Just in time (JIT) compiler and automatic garbage collection to perform the critical task [15].

For transparent communication between Java Virtual Machines, it has Remote Method Invocation (RMI). It is utilized to develop high performance computing applications.

### 5.8. High Performance FORTRAN (HPF)

Its name conveys that it is the extension of Fortran 90. It supports the parallel programming principles [16]. It supports the data parallel programming pattern in which one program has the complete control for the distribution of data among all the processors. It works in the distributed memory environment. It is a portable programming language.

### 5.9. Z-level Programming Language (ZPL)

It is a language with parallelized compiler. It is particularly for the high performance computations such as the scientific as well as engineering. It abstracts the Flynn's MIMD (Multiple Instructions and Multiple Data) parallel architecture [17]. The applications developed in this language are portable and the performance is independent of the compiler as well as the machine. It is a good programming language, but the scientists and the engineers have not shown much interest in it.

### 5.10. Erlang

It is a functional programming language. *Firstly*, it was introduced by the telecommunication giant Ericsson to build the telecommunication switches. Lately in 1998, it becomes open source software [18]. Concurrency is achieved through threads. The applications developed in this language are highly available as well as reliable. In this programming paradigm the explicit threading parallel mechanism is utilized in which the programmers create the explicit threads to achieve the parallelism [23].

### 5.11. Unified Parallel C (UPC)

It supports both types of architectures shared memory as well as distributed memory. It is based on the partitioned memory principle [12]. The complete memory is partitioned into many small memory areas for every thread. Every thread has a private memory as well as global memory which are shared among the same class of threads. A new principle is used to get high performance that is thread affinity in which memory access performance among the threads of same class is optimized [19]. Workload management in it is implied and the work partitioning as well as workers mapping may be implied or programmer controlled. The thread communication is maintained with the help of pointers. Three types of pointers are utilized here which are as follows: (i) the private pointers who works on their own address spaces, (ii) sharing pointer who works on the shared memory area, and (iii) sharing pointers to share, these are the sharing pointers who works on the other shared memory. So many synchronization mechanisms are utilized in this language like barrier, split phase barriers, fence, locks, and memory consistency control. It resemble with the MPI platform in workload partitioning and worker mapping.

### 5.12. Streams and Iteration in a Single Assignment Language (SISAL)

It is a functional programming language. It offers automatic parallelism through its functional semantics. In it user-defined names are identifiers rather than variables. These identifiers are known as values rather than the memory locations [20]. These values are dynamic entities. The identifiers are defined and bound to the values only for the time of execution. The Sisal Compiler is optimizing one which converts the source program into the object code with the execution time system parts needed to automatic managing of memory, tasks, and input/output. The parallelism can be controlled by the users also. In conclusion the programming language has an optimizing compiler with better runtime performance.

### 5.13. Laboratory Virtual Instrumentation Engineering Workbench (*LabVIEW*)

It is visual programming language from National Instruments. It is a platform as well as development environment. It is also a data-flow programming language in which the execution is decided with the help of structure of a graphical block diagram by drawing the wires for the function nodes [21].These wires propagate the variables and the nodes starts to execute as soon as the input data is available.  This programming language is basically utilized for the acquiring data and processing signals, instrument control, automating test and validation systems, and monitoring and controlling embedded systems. It may be run on a number of platforms like MS Windows, UNIX, Linux, and Mac OS X. Multiprocessing as well as multi-threaded hardware are automatically utilized by its inbuilt schedulers. We can also create distributed applications on this platform. Hence it is a good high performance computing technology. The non-programmers can also develop the good applications by dragging and dropping the virtual representations of the laboratory equipments to whom they are well-known.

### 5.14. Manticore Programming language

It is a new functional parallel programming language. It is a heterogeneous programming language that provides the parallelism at multiple levels. It provides coarse-grained, explicit parallelism based on Concurrent ML platform. It supports the explicit concurrency with fine-grain and implicit threads [22]. The synchronization is provided with first class synchronization message passing which well fits to the nature of the functional programming paradigms [23]. The locally-concurrent/globally-sequential garbage collector is implemented.

## 6. CONCLUSIONS

In this construct, a little survey of parallel programming languages, their design approaches, and their constructs are presented. The current scenario is totally towards the parallelism to achieve the high performance computing (HPC) and the developers must be aware about the new concepts of the technology. And this article is good food for the novice parallel programming lovers who want to do much more in this field.